\documentclass[letterpaper, 10 pt, conference]{ieeeconf}  % Comment this line out if you need a4paper

\IEEEoverridecommandlockouts                              % This command is only needed if 
\usepackage{graphicx}
\usepackage{amsmath}
\usepackage{cite}
\usepackage{tabulary}
\usepackage{amssymb}
\usepackage{subcaption}

\usepackage{multirow}
\usepackage{booktabs}

\title{\LARGE \bf
Suitability of an inter-burst detection method for grading
hypoxic-ischemic encephalopathy in newborn EEG}

\author{Sumit A. Raurale$^{1,2}$ \textit{Student Member, IEEE}, Saif Nalband$^{1,2}$,
  Geraldine B. Boylan$^{1,2}$, \\Gordon Lightbody$^{1,3}$, and John M.
  O'Toole$^{1,2} \textit{Member, IEEE}$ % <-this % stops a
  \thanks{*This work is supported by the Wellcome Trust (209325/Z/17/Z) and Science Foundation
    Ireland (INFANT-12/RC/2272).  
    JMOT is supported by Science Foundation Ireland (15/SIRG/3580 and
    18/TIDA/6166).}% <-this
  \thanks{$^{1}$Irish Centre for Fetal \& Neonatal Translational Research (INFANT),
    University College Cork, Ireland. 
    {\tt\small (sumit.raurale@ucc.ie)}}%
  \thanks{$^{2}$Department of Paediatrics and Child Health, University College Cork,
    Ireland.}%
  \thanks{$^{3}$Department of Electrical \& Electronic Engineering, University College
    Cork, Ireland.}%
}

\begin{document}

\maketitle
\thispagestyle{empty}
\pagestyle{empty}

%%%%%%%%%%%%%%%%%%%%%%%%%%%%%%%%%%%%%%%%%%%%%%%%%%%%%%%%%%%%%%%%%%%%%%%%%%%%%%%%
\begin{abstract}
  % Electroencephalography (EEG) is an important clinical tool to guide treatment decision on term infants based on background period as inter-bursts interval (IBI) activity to detect hypoxic-ischemic encephalopathy (HIE) abnormalities. This paper validates an inter-burst detection method developed for premature infants EEG to differentiate inter-burst activity on term infants which is used for grading HIE abnormalities. It is shown how simple IBI statistical measure with MLP classifier can successfully identify four major HIE abnormalities in term infants with overall 77.8\% classification accuracy. The proposed system is the first to use such inter-burst detection approach targeted for premature infants to successively grade four major HIE abnormalities in term infants EEG.

Electroencephalography (EEG) is an important clinical tool for grading injury caused by lack of oxygen or blood to the brain during birth. Characteristics of low-voltage waveforms, known as inter-bursts, are related to different grades of injury. This study assesses the suitability of an existing inter-burst detection method, developed from preterm infants born $<$30 weeks of gestational age, to detect inter-bursts in term infants. Different features from the temporal organisation of the inter-bursts are combined using a multi-layer perceptron (MLP) machine learning algorithm to classify four grades of injury in the EEG. We find that the best performing feature, percentage of inter-bursts, has an accuracy of 59.3\%. Combining this with the maximum duration of inter-bursts in the MLP produces a testing accuracy of 77.8\%, with similar performance to existing multi-feature methods. These results validate the use of the preterm detection method in term EEG and show how simple measures of the inter-burst interval can be used to classify different grades of injury.  
\end{abstract}

%%%%%%%%%%%%%%%%%%%%%%%%%%%%%%%%%%%%%%%%%%%%%%%%%%%%%%%%%%%%%%%%%%%%%%%%%%%%%%

%Discontinuous EEG pattern observed in infants consist of low voltage activity followed by high voltage activity known as inter-burst and burst respectively \cite{R2}.

\section{Introduction}
\label{introduction}
Electroencephalography (EEG) provides an effective non-invasive tool for detecting abnormalities of the brain. Different EEG activity can provide clinical information with respect to changes in sleep \cite{R3}, maturation \cite{R5}, neuro-developments \cite{R4}, and hypoxic-ischemic encephalopathy (HIE) \cite{R1}, a type of injury caused by lack of blood and oxygen to the brain. In the clinical care of asphyxiated infants, it is essential to correctly identify the degree of HIE severity from EEG recording \cite{R1}. But continuous EEG monitoring and review in most neonatal intensive care units is not practical. Automated EEG analysis could help with this review by presenting the physician with useful and timely clinical information about brain function.  

A distinct marker of EEG abnormalities, and therefore HIE is a discontinuous pattern in the EEG of term infants. This activity consists of short-duration periods of high-voltage activity, known as bursts, interspersed with periods of low voltage activity, known as inter-bursts \cite{R2}. From literature, not many studies have been carried on detecting brain abnormalities in term infants through EEG analysis using inter-burst detection or features based on inter-burst activity \cite{R6}. Amongst such is a study by O'Toole \emph{et al.} which has developed an inter-burst detection approach for preterm infants using multi-feature and machine learning approach \cite{R7}. Besides this, there are few studies which shows burst detection among term infants. Matic \emph{et al.} \cite{R6} has used adaptive inter-bursts interval detection by extracting six significant features from the burst interval activities followed by pattern classification. Parisa \emph{et al.} \cite{R8} have recently used non-linear feature extraction techniques with machine learning classification for burst detection. These studies show significant results for detecting burst or inter-burst activity, on preterm or term infants EEG.
%Also, none of the study have shown validation of preterm infants inter-burst detection approach on term infants.

An integral part of grading the EEG for HIE injury, is to assess the level of discontinuity in the EEG by visually quantifying inter-burst activity \cite{R1,R9}. Automated grading methods developed by Stevenson \emph{et al}. \cite{R11} and Ahmed \emph{et al}. \cite{R12} which use multiple complex time-domain, frequency-domain, and information theory features. These studies do not consider detailed classification based on inter-burst interval analysis, which is an important component of visual grading of the EEG.

Based on the above shortcomings, this paper provides the following contributions:
\begin{itemize}
\item Validate an inter-burst detection method developed for preterm EEG \cite{R7}, to investigate its ability to detect inter-burst intervals (IBIs) on the EEG of term infants.
\item Differentiate the four grades of HIE using an automated EEG system with simple features of the inter-bursts measure followed by a data-driven MLP approach.
\end{itemize}

%But, distinguishing between burst and inter-burst activities within EEG for pre-term and term infants is an important aspect of analysis for detecting brain abnormalities at early stage.

%The remainder of this paper is as follows. Section 2 describes the Inter-Burst detection approach for pre-term neonates EEG signal, before Section 3 presents our concept approach for grading EEG abnormality in term neonates, the performance of which is analysed in Section 4.

\section{Methods}
\begin{figure*}[h]
	\centering
	\vspace{0.15cm}
	\includegraphics[width=0.96\linewidth]{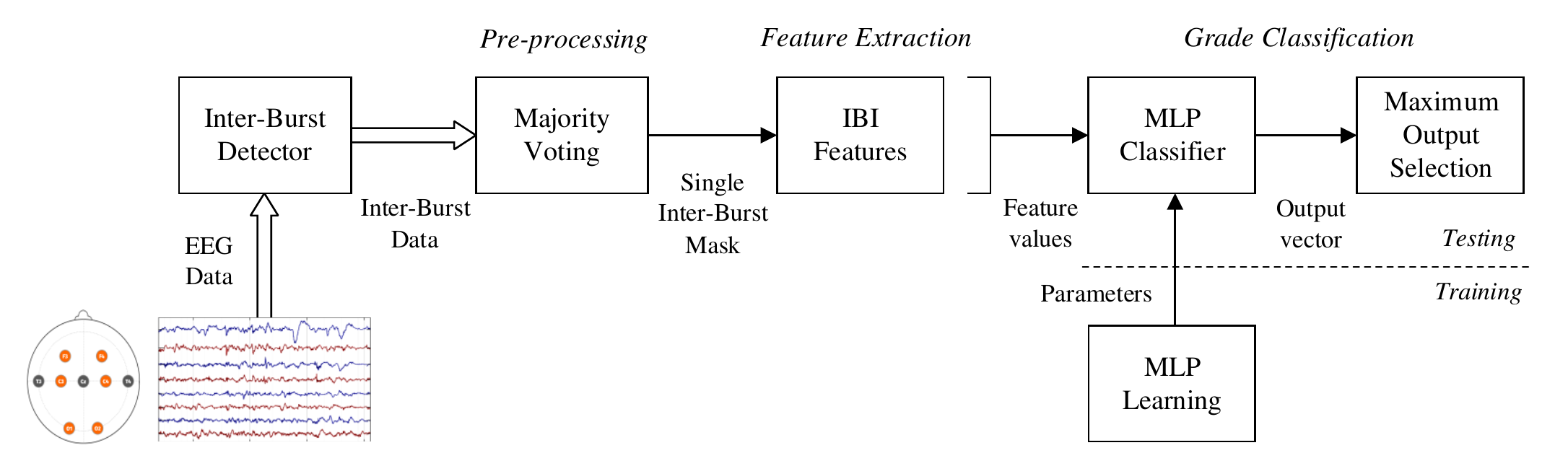}
	%\vspace{0.15cm}
	\caption{Proposed HIE grading system for term infant EEG. (Majority voting: multiplexing data into individual binary masks through majority count.)}
	\label{system}
\end{figure*}

%We aim to develop an automated system to classify the grade as severity of HIE from a 1-hour EEG epoch recorded from term infants.
We implement a burst detection method developed for preterm infant EEG \cite{R7} and investigate its ability to detect IBIs on the EEG of term infants. The detector classifies the EEG as bursts or inter-bursts, but for our application we are only interested in characteristics of the inter-bursts. We then extract features from the temporal organisation of the IBIs and use a data-driven approach to differentiate four grades of HIE. The overall structure of the proposed system is shown in Fig. \ref{system}.

%Need to explain what the epoch is (and are you still using it?). Also, what about summarising across channels? And should describe all acronyms used in figures.

%[COMMENTS: 1) can you explain all acronyms in the caption; 2) 'bilateral inter-burst' is summary across all channels? if so then better to say average (majority vote) over all channels; 3) replace 'Grade Detection' with 'Grade classification'; 4) 'Feature set': shouldn't have decided at this stage that only using 2 features; maybe better to replace with generic 'IBI features' or something.]

\subsection{EEG Data}

The EEG was recorded on 54 term infants using a NicoletOne EEG system within the neonatal intensive care unit (NICU) of Cork University Maternity Hospital, Cork, Ireland. This study was approved from the Clinical Ethics Committee of the Cork Teaching Hospital with written and informed parental consent obtained before EEG recording. The EEG recordings was initiated within 6 hours of birth and continued for up to 72 hours to monitor the evolution of the developing encephalopathy with seizure surveillance. The EEG recording used 9 active electrodes T4, T3, O1, O2, F4, F3, C4, C3, and Cz as standard protocol in the NICU. Our analysis used an 8-channel bipolar montage derived from these electrodes as F4-C4, C4-O2, F3-C3, C3-O1, T4-C4, C4-Cz, Cz-C3 and C3-T3. 

One-hour epochs of EEG were pruned from the continuous EEG to avoid major artefacts when possible. This epoch was reviewed independently by 2 EEG experts and graded according to the system defined by Murray \emph{et al.} \cite{R1}, described in Table \ref{grading}. If the experts disagreed, they reviewed the EEG together and a consensus was reached to determine the final grade. This same dataset has been used by Stevenson \emph{et al}. \cite{R11} and Ahmed \emph{et al}. \cite{R12}.  

\begin{table}[h]
\renewcommand{\arraystretch}{1.30}
\centering
\caption{EEG grades for hypoxic--ischemic encephalopathy defined by Murray \emph{et al}. \cite{R1}.}
\label{grading}
\begin{tabular}{c@{\hskip 0.25cm}l}
\toprule
\textbf{Grade} & \textbf{Description of EEG} \\
\midrule
1 & Normal/Mild abnormalities: Continuous background pattern\\
& with mild voltage depression and asymmetric patterns\\
2 & Moderate abnormalities: Discontinuous activity with clear\\
& asymmetry, IBI $\leq$ 10 s\\
3 & Major abnormalities: Discontinuous activity with IBI range\\
& between 10-60 s, severe depression without wake/sleep cycles\\
4 & Inactive/Severe abnormalities: Background activity pattern\\
&  $\leq$ 10$\mu$V or severe discontinuity of IBI $\geq$ 60 s\\
\bottomrule
\end{tabular}
\end{table}

\subsection{Inter-Burst Detection Approach}

\begin{figure}[h]
\centering
\begin{subfigure}[b]{0.45\textwidth}
   \includegraphics[width=1\linewidth]{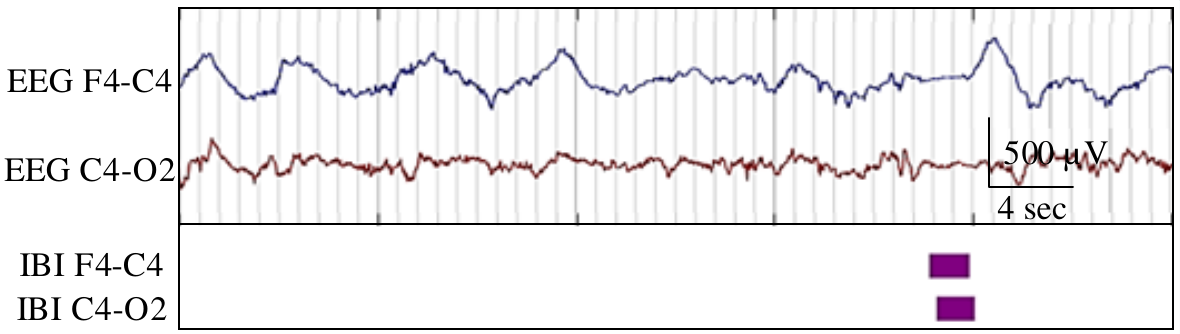}
   \vspace{-0.45cm}
   \caption{Normal/Mild abnormalities}
   \label{fig:Ng1} 
   \vspace{0.3cm}
\end{subfigure}
\vspace{0.2cm}
\begin{subfigure}[b]{0.45\textwidth}
   \includegraphics[width=1\linewidth]{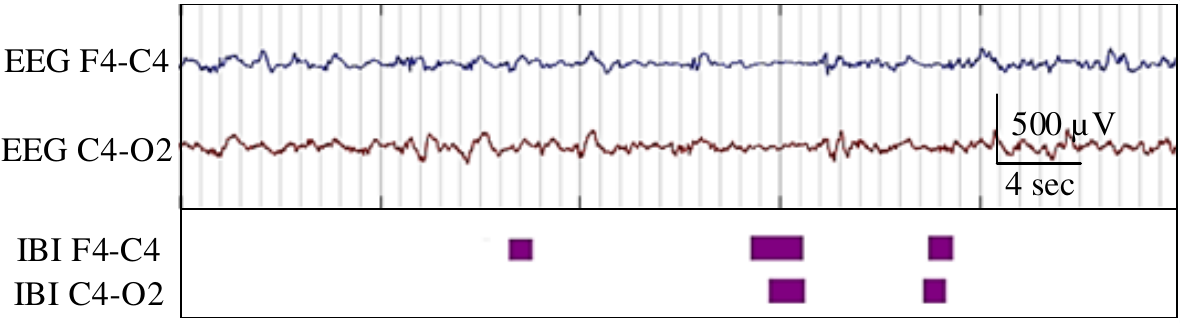}
   \vspace{-0.45cm}
   \caption{Moderate abnormalities}
   \label{fig:Ng2}
   \end{subfigure}
   \vspace{0.3cm}
\begin{subfigure}[b]{0.45\textwidth}
   \includegraphics[width=1\linewidth]{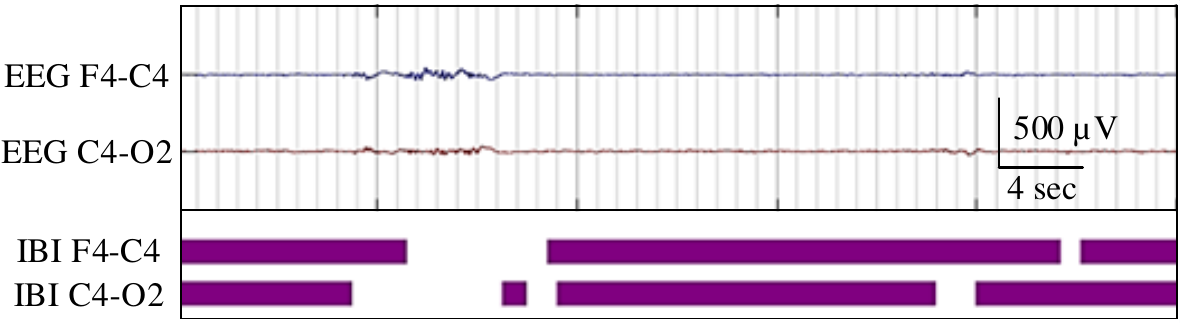}
   \vspace{-0.45cm}
   \caption{Major abnormalities}
   \label{fig:Ng3}
\end{subfigure}
\vspace{0.1cm}
\begin{subfigure}[b]{0.45\textwidth}
   \includegraphics[width=1\linewidth]{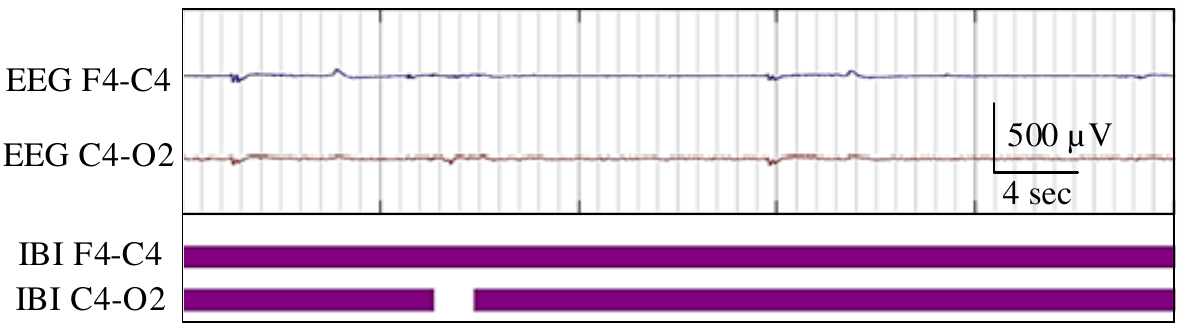}
   \vspace{-0.45cm}
   \caption{Severe abnormalities}
   \label{fig:Ng4}
\end{subfigure}
\caption{Detecting inter-burst intervals (IBI) for 4 different EEG grades.}
\label{fig_sim}
\end{figure}

In our proposed system architecture, we incorporate an inter-burst detection method which was initially designed for preterm EEG. The method extracts multiple features of amplitude, spectral shape and power, and a frequency-weighted energy features \cite{R7}. These features are estimated within four frequency bands 0.5--3, 3--8, 8--15, and 15--30~Hz. A feature selection method is used to maximise relevancy and minimise redundancy, retaining only necessary features. These selected features were then combined using a linear-kernel support vector machine (SVM). This channel-independent system was trained on EEG data from 36 premature infants with a gestational age of $<$ 30 weeks. We test this method on term EEG, processing each channel separately to detect inter-burst activity. Visual assessment of the method demonstrated reasonable performance at detecting inter-bursts, as the example in Fig.~\ref{fig_sim} shows for different grades.

% Unlike existing methods, which either operate on 1 specific channel [17] or all channels simultaneously [6,18] ,

\subsection{Preprocessing and Feature Extraction}
Each bipolar EEG channel was analysed by the inter-burst detector. As HIE is consider a global brain injury \cite{R10}, the inter-burst information is compressed over the 8 channels by majority voting of the individual binary masks to form one summary mask. We then extract two features to summarise the binary mask.

\vspace{0.25cm}
\subsubsection{IBI percentage}
\label{sssec:iper}
IBI percentage ($IBI\%$) is used to detect the overall occurrence of inter-burst activity in an epoch. It is calculated as the summation of duration of the IBIs to the overall epoch length,

\begin{equation}
IBI\% = \dfrac{\sum_i IBI_i}{L}\times 100
\label{equ1}
\end{equation}

\noindent where, $IBI_i$ represents the duration of $i$-th IBI and $L$ denotes the overall length of epoch.

\vspace{0.25cm}
\noindent \subsubsection{Maximum IBI}
\label{sssec:lint}
The HIE grading describes various maximum IBIs for different grades. We calculate this measure by subtracting the difference between the maximum and the minimum inter-burst length within an epoch as,
\begin{equation}
IBI_m = \max (IBI_{i})-\min (IBI_{i}).  
\label{equ2}
\end{equation}

\vspace{0.10cm} \subsection{Classification of HIE Grades}
To combine features and to apply a data-driven approach to estimating thresholds for the 4 class separation, we use a multi-layer perceptron (MLP) as the machine learning algorithm.  
% After extracting the features from the epoch, there still some redundancy between classes
% because of hard thresholds for the different grades which makes it difficult to evaluate a
% grading decision from the feature itself. Thus, for identifying the uncommon states
% throughout the varying feature values, we use a data-driven approach using multi-layer
% perceptron (MLP) which set its own thresholds by combining the feature values.

The MLP was configured with one hidden layer with fourteen neurons and four neurons in output layer, one for each HIE grade to be recognised. The selection criterion was based on the convergence of the learning error from different combinations as described in \cite{R13}. Weights and biases were initialised before training from a normal distribution with a mean and variance of 0 and 1, respectively. The training process was stopped when the absolute rate of change in the average squared error ($\theta$) per iteration was sufficiently small ($\theta \leq 0.1$) \cite{R13}. The MLP model was trained based on the evaluated feature dataset and while testing, the maximum value within the output layer neuron in MLP was selected as the recognised grade for an evaluated feature set from epoch.

For testing we use a leave-one-subject-out (LOSO) cross-validation method. Based on 54 infants considered in study, system performance was observed by training the MLP classifier with 53 EEGs and testing on the one left out. This was progressed until all 54 infant's EEGs were tested to determine the overall classification accuracy of system. 
\vspace{0.25cm}
\section{Results}
\label{expresult}
\begin{figure}[h!]
\centering
\begin{subfigure}[b]{0.39\textwidth}
   \includegraphics[width=1\linewidth]{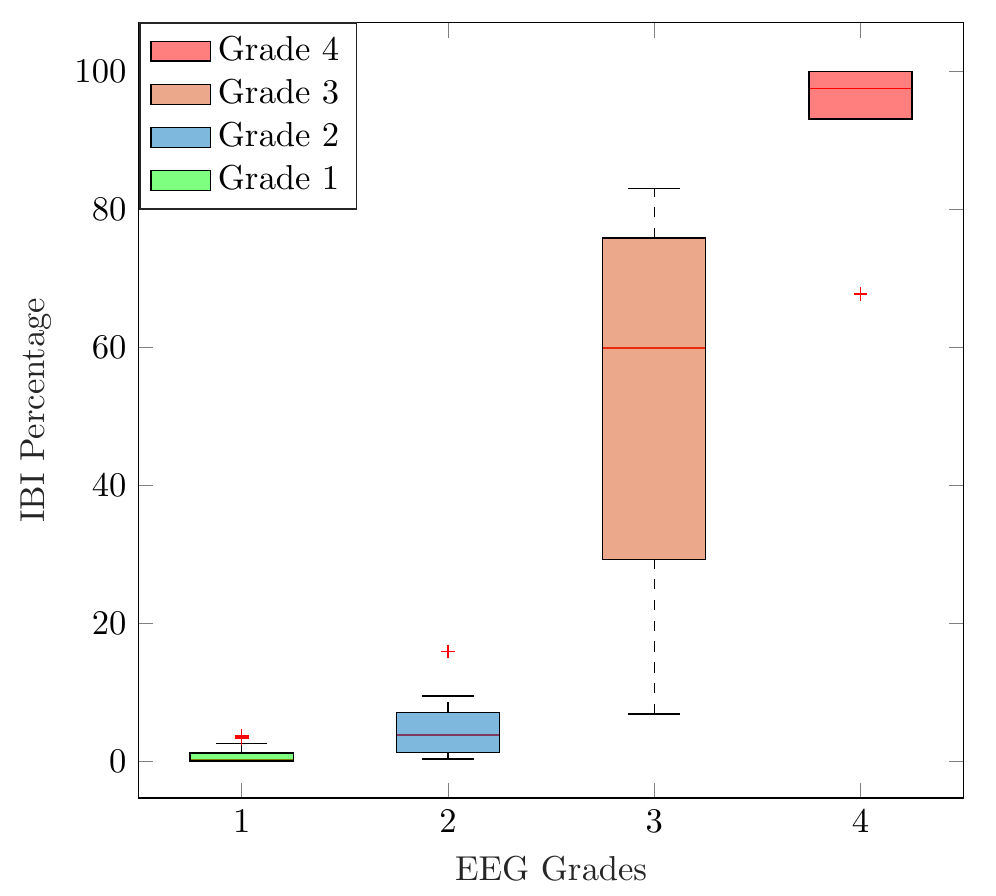}
   \vspace{-0.45cm}
   \label{fig:Var1} 
   \vspace{0.5cm}
\end{subfigure}
%\vspace{0.2cm}
\begin{subfigure}[b]{0.395\textwidth}
   \includegraphics[width=1\linewidth]{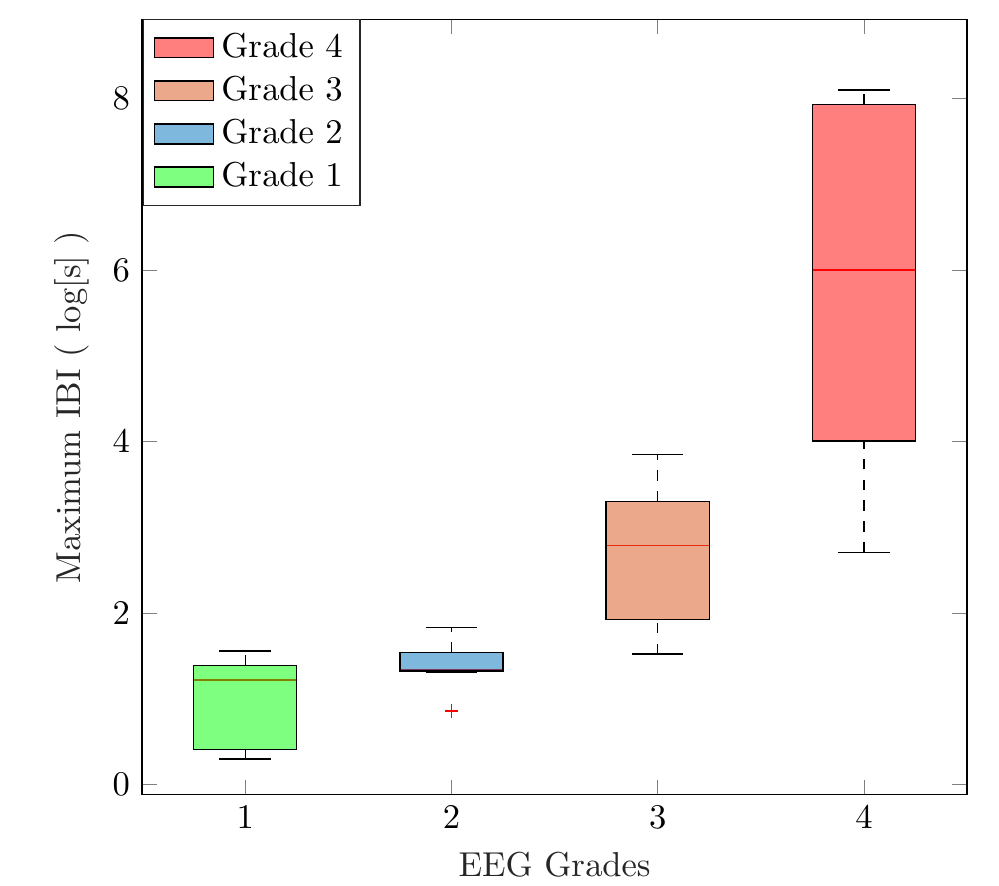}
   %\vspace{-0.45cm}
   \label{fig:Var2}
   \end{subfigure}
\caption{Distribution of the IBI features separated by EEG grade (IBI: inter-burst interval). The maximum IBI feature is plotted on log scale which represents 2.3 for 10 s and 4.1 for 60 s thresholds of the grading system in Table \ref{grading}. Box represent inter-quartile range, red lines represent median values, and error bars represent 95th centiles.}
\label{feat_var}
\end{figure}

\vspace{0.10cm}
Fig.~\ref{feat_var} illustrates distribution of the IBI percentage and maximum IBI features for each of the 4 EEG grades. We find varying degrees of separation between the 4 classes for the different features.  

The evaluated IBI percentage shows the lowest percentage of IBI activity for Grade 1 followed by Grade 2 and nearly 30--80\% for Grade 3 after Grade 4 showing close to 100\% of IBI activity which satisfy the clinical EEG grading condition for HIE defined by Murray \emph{et al}. Further, the estimated maximum IBI duration for Grade 1 and Grade 2 falls under 10 s, Grade 3 range approximately between 10 -- 60 s and the maximum IBI duration for Grade 4 mostly exceeds 60 s which strongly correlates to the clinical grading system defined by Murray \emph{et al.} in Table \ref{grading}.

Testing results using each IBI feature separately is $<$ 60\% accuracy, as shown in Table~\ref{fsaccuracy}. Combining both features, however, greatly improves performance to 77.8\%.

\begin{table}[h]
\renewcommand{\arraystretch}{1.35}
\centering
\caption{Performance of the multi-layer perceptron using 1 and 2 features.}
\label{fsaccuracy}
\begin{tabular}{c | c}
\hline 
\hline 
\textbf{Features} & \textbf{Classification accuracy} \\
\hline
IBI percentage & 59.3\% \\
maximum IBI & 53.7\% \\
IBI percentage \& maximum IBI & 77.8\% \\
\hline
\hline
\end{tabular}
\end{table}

Confusion matrix for the LOSO cross-validation using the two features is presented in Table \ref{confusion}. It shows that 12 out of 54 infant data were misclassified resulting in the accuracy of 77.8\%. Most misclassification occurred between grades 1 and 2, which from Fig.~\ref{fig_sim} appears to be the harder grades to separate using these IBI features alone.  

\begin{table}[h]
\renewcommand{\arraystretch}{1.45}
\centering
\vspace{0.30cm}
\caption{Confusion matrix of the system's output and actual grade}
\label{confusion}
\setlength\tabcolsep{10.0pt}
\begin{tabular}{c c c c c c c}
\hline 
Actual& \multicolumn{4}{c}{System's Output} & Total & Incorrect\\
 \cline{2-5}
grade & 1 & 2 & 3 & 4 & &  \\ 
\hline
1 & {\bf 20} & 2 & 0 & 0 & 22 & 2\\
2 & 4 & {\bf 8} & 2 & 0 & 14 & 6\\
3 & 1 & 2 & {\bf 9} & 0 & 12 & 3\\
4 & 0 & 0 & 1 & {\bf 5} & 6  & 1\\
\hline
Precision & 25 & 12 & 12 & 5 & 54 & 12\\
\hline
\end{tabular}
\end{table}

The proposed system performance is compared with other approaches which uses the same EEG dataset, as presented in Table \ref{compare}. Stevenson \emph{et al}. \cite{R11} shows similar 77.8\% classification accuracy using a complex feature set of 8 values. Ahmed \emph{at al}. \cite{R12} achieve a higher level of accuracy (87\%) using Gaussian mixture model supervector combined with an SVM with a larger feature set of 55 values from time, frequency, and information theory features, in addition to probabilistic post-processing analysis. In contrast, our proposed system using only 2 features of the IBI measure, manages comparable accuracy to the Stevenson \emph{et al}. method.  
% burst-detection method developed for premature EEG, with two simple statistical features
% and MLP data-driven approach which takes the decision from its own IBI set thresholds.

%Also, none of these studies have considered the classification approach based on defined IBI parameters for HIE abnormalities \cite{R1}, which could be unreliable to use in real-life practice. 

\begin{table}[h]
\renewcommand{\arraystretch}{1.45}
\centering
\caption{Comparison of the proposed method with the other techniques.}
\label{compare}
\setlength\tabcolsep{10.0pt}
\begin{tabular}{c | c c c}
\toprule
HIE Grading & Stevenson & Ahmed & Proposed \\
Method & \textit{et al}. \cite{R11} & \textit{et al}. \cite{R12} & method \\
\midrule
Overall accuracy & 77.8\% & 87.0\% & 77.8\% \\ 
\bottomrule
\end{tabular}
\end{table}

%Moreover, by training the inter-burst detection algorithm with term infants EEG data, and adding more statical features based on IBI measure in accordance to HIE grading criteria, the proposed system performance can be well improved.

\section{Discussion and Conclusions}
\label{discussion}
\vspace{0.1cm}
We present a novel approach for grading HIE abnormalities in the EEG. The
results validate the use of a preterm inter-burst detection method to detect inter-bursts in the EEG of term newborns. Simple measures of the IBI combined using a data-driven approach can achieve comparable accuracy with state-of-the-art HIE grading systems.

A previous study by Stevenson \emph{et al}. \cite{R11} reported an accuracy of 77.8\% while testing on same dataset of 54 term infants. This performance improved to 87\% by Ahmed \emph{et al}. using a large feature set, GMM supervectors combined with SVM, and additional post-processing \cite{R12}. Both these studies use complex multi feature analysis on each epoch without considering the classification of inter-burst interval based on the clinical  criterion. Future work will include developing a more complete feature set to capture other characteristics of the EEG to further improve classification performance.

In conclusion, we validate the use of an EEG inter-burst detection method developed for preterm infants applied to the EEG of term infants. Only two simple features of the temporal organisation of the inter-bursts are used to generate a high-performance system comparable in accuracy with more complex multi-feature methods \cite{R11}. Future systems will incorporate these IBI features with a more diverse feature set and optimised machine learning methods to further improve performance.  

% , and a data-driven
% approach using MLP which set its own thresholds by combining the temporal inter-bursts
% values even from the hard thresholds for the different grades. The developed automated HIE
% grading method shows similar performance to an existing, complex multi-feature method by
% Stevenson \textit{et al}. \cite{R11}. Also, have successively validated the use
% inter-burst detection method for premature infants EEG with the term infants.


\vspace{0.20cm}
\begin{thebibliography}{99}
\vspace{0.15cm}
\bibitem{R3} P. Kirsi, T. Kirjavainen, S. Stjerna, T. Salokivi, and S. Vanhatalo. ``Sleep wake cycling in early preterm infants: comparison of polysomnographic recordings with a novel EEG-based index" \textit{Clinical Neurophysiology}, 124(9), (2013): 1807--1814.
\vspace{0.1cm}
\bibitem{R5} J. M. O'Toole, G. B. Boylan, S. Vanhatalo, and N. J. Stevenson. ``Estimating functional brain maturity in very and extremely preterm neonates using automated analysis of the electroencephalogram." \textit{Clinical Neurophysiology}, 127(8), (2016): 2910--2918.
\vspace{0.1cm}
\bibitem{R4} R. W. Claire, J. E. Harding, C. E. Williams, M. I. Gunning, and M. R. Battin. ``Quantitative electroencephalographic patterns in normal preterm infants over the first week after birth" \textit{Early Human Development}, 82(1), (2006): 43--51.
\vspace{0.1cm}
\bibitem{R1} M. D. Murray, C. A. Ryan, G. B. Boylan, A. P. Fitzgerald, and S. Connolly. ``Prediction of seizures in asphyxiated neonates: correlation with continuous video-electroencephalographic monitoring" \textit{Pediatrics}, 118(1), (2006): 41--46.
\vspace{0.1cm}
\bibitem{R2} S. Vanhatalo, J. M. Palva, S. Andersson, C. Rivera, J. Voipio, and K. Kaila. ``Slow endogenous activity transients and developmental expression of K+-Cl- cotransporter 2 in the immature human cortex" \textit{European Journal of Neuroscience}, 22(11), (2005): 2799--2804.
\vspace{0.1cm}
\bibitem{R6} V. Matic, P. J. Cherian, K. Jansen, N. Koolen, G. Naulaers, R. M. Swarte, P. Govaert, S. V. Huffel, and M. D. Vos. ``Improving reliability of monitoring background EEG dynamics in asphyxiated infants." \textit{IEEE Transactions on Biomedical Engineering}, 63(5), (2016): 973--983.
\vspace{0.1cm}
\bibitem{R7} J. M. O'Toole, G. B. Boylan, R. O. Lloyd, R. M. Goulding, S. Vanhatalo, and N. J. Stevenson. ``Detecting bursts in the EEG of very and extremely premature infants using a multi-feature approach." \textit{Medical Engineering \& Physics}, 45, (2017): 42--50.
\vspace{0.1cm}
\bibitem{R8} P. Mirzaei, G. Azemi, N. Japaridze, and B. Boashash. ``Surrogate data test for nonlinearity of EEG signals: A newborn EEG burst suppression case study." \textit{Digital Signal Processing}, 70 (2017): 30--38.
\vspace{0.1cm}
\bibitem{R9} A. Piryatinska, T. Gyorgy, W. A. Woyczynski, K. A. Loparo, M. S. Scher, and A. Zlotnik. ``Automated detection of neonate EEG sleep stages." \textit{Computer Methods and Programs in Biomedicine}, 95(1), (2009): 31--46.
\vspace{0.1cm}
\bibitem{R11} N. J. Stevenson, I. Korotchikova, A. Temko, G. Lightbody, W. P. Marnane, and G. B. Boylan. ``An automated system for grading EEG abnormality in term neonates with hypoxic-ischaemic encephalopathy." \textit{Annals of Biomedical Engineering}, 41(4), (2013): 775--785.
\vspace{0.1cm}
\bibitem{R12} R. Ahmed, A. Temko, W. Marnane, G. Lightbody, and G. Boylan. ``Grading hypoxic–ischemic encephalopathy severity in neonatal EEG using GMM supervectors and the support vector machine." \textit{Clinical Neurophysiology}, 127(1), (2016): 297-309.
\vspace{0.1cm}
\bibitem{R10} B. H. Walsh, D. M. Murray, and G. B. Boylan. ``The use of conventional EEG for the assessment of hypoxic ischaemic encephalopathy in the newborn: a review." \textit{Clinical Neurophysiology}, 122(7), (2011): 1284--1294.
\vspace{0.1cm}
\bibitem{R13} S. Raurale, J. McAllister, and J. M. del Rincon, ``EMG acquisition and hand pose classification for bionic hands from randomly-placed sensors" in \textit{2018 IEEE International Conference on Acoustics, Speech and Signal Processing (ICASSP)}, (2018): 1105--1109.

\end{thebibliography}
\end{document}